\documentclass[aps,prl,twocolumn,groupedaddress,nofootinbib,showpacs]{revtex4}
\usepackage[dvips]{graphicx}
\usepackage{color}
\usepackage{cancel}
\usepackage{amsfonts}
\usepackage{dsfont}
\newif\ifdraft
\drafttrue
\newif\ifpreprint
\preprinttrue

\def\id{\protect{{1 \kern-.28em {\rm l}}}}

\def\spa#1.#2{\left\langle#1\,#2\right\rangle}
\def\spb#1.#2{\left[#1\,#2\right]}

\def\Tr{{\rm Tr}}

\def\gR{g_{\text{R}}}
\newcommand{\bea}{\begin{eqnarray}}
\newcommand{\eea}{\end{eqnarray}}
\newcommand{\ee}{\end{equation}}
\newcommand{\be}{\begin{equation}}
\newcommand{\eq}{\begin{equation}}
\newcommand{\eqe}{\end{equation}}
\newcommand{\eqa}{\begin{eqnarray}}
\newcommand{\eqae}{\end{eqnarray}}

\newbox\charbox
\newbox\slabox
\def\s#1{{      
        \setbox\charbox=\hbox{$#1$}
        \setbox\slabox=\hbox{$/$}
        \dimen\charbox=\ht\slabox
        \advance\dimen\charbox by -\dp\slabox
        \advance\dimen\charbox by -\ht\charbox
        \advance\dimen\charbox by \dp\charbox
        \divide\dimen\charbox by 2
        \raise-\dimen\charbox\hbox to \wd\charbox{\hss/\hss}
        \llap{$#1$}
}}

\def\no{\nonumber}
\def\cN{{\cal N}}
\def\cR{{\cal R}}
\def\cG{{\cal G}}

\begin{document}

\title{
\ifpreprint
\hbox{\rm \small
UUITP-37/17
\hskip 2.8 cm
IGC-17/10-2
\hskip 2.8 cm
NORDITA 2017-110
\hskip 2.8 cm
NSF-ITP-17-137
\break}

\fi
\vskip 0.1cm
Gauged supergravities and spontaneous SUSY breaking from the double copy}
\vskip -0.2cm
 
\author{M.~Chiodaroli,${}^{a,d}$ M.~G\"{u}naydin,${}^{b,d}$ H.~Johansson,${}^{a,c,d}$ and R.~Roiban${}^{b,d}$}

\affiliation{\vskip 0.1cm
${}^a$Department of Physics and Astronomy, Uppsala University, 75108 Uppsala, Sweden\\
${}^b$Institute for Gravitation and the Cosmos, The Pennsylvania State University, University Park PA 16802, USA \\
${}^c$Nordita, Stockholm University and KTH Royal Institute of Technology, Roslagstullsbacken 23, 10691 Stockholm, Sweden \\
${}^d$Kavli Institute for Theoretical Physics, University of California, Santa Barbara, CA 93106, USA}

\begin{abstract} 

Supergravities with gauged R-symmetry and Minkowski vacua allow for spontaneous supersymmetry breaking and, as such, provide a 
framework for building supergravity models of phenomenological relevance.  
In this letter we initiate the study of double-copy constructions for these  supergravities. 
We argue that, on general grounds, we expect their scattering amplitudes to be described by a double copy of the type: (Higgsed gauge theory)$\,\otimes\,$(gauge theory with broken SUSY).   
We  present a simple realization in which the resulting supergravity has $U(1)_{\text{R}}$ gauge symmetry, spontaneously-broken $\cN=2$ supersymmetry, and massive gravitini.
This is the first instance of a double-copy construction of a gauged supergravity and of a theory with spontaneously-broken supersymmetry. The construction extends in a 
straightforward manner to a large family of gauged Yang-Mills-Einstein supergravity theories with or without 
spontaneous gauge-symmetry breaking.

\end{abstract}

 \pacs{04.65.+e, 11.15.Bt, 11.30.Pb, 11.55.Bq \hspace{1cm}} 

\maketitle

Gauged supergravities---supergravities in which part 
of the R-symmetry has been promoted to a gauge symmetry---play a central role in the diverse landscape of supersymmetric extensions of gravity.
From a string-theory perspective, such theories arise naturally in the context of flux compactifications. Certain classes of gauged supergravities admit flat Minkowski vacua in which supersymmetry is (spontaneously) broken and hence can be employed in the search for phenomenologically-viable models. 
Classifying all possible supergravity gaugings in various dimensions has been 
the object of a large body of work (see reviews \cite{Salam:1990ey, Freedman:2012zz}), eventually leading to the formulation of the embedding-tensor formalism \cite{Nicolai:2001sv,Nicolai:2000sc,deWit:2002vt,deWit:2005hv,deWit:2004nw} (see also~\cite{SamtlebenTrigianteLectures}) that resulted in the discovery of novel gaugings including a new family of $SO(8)$ gauged supergravities in four dimensions \cite{DallAgata:2012mfj}.

Recent advances in scattering amplitude calculations have been playing a key role in revealing hidden properties of gravity. Amplitudes in many supergravities admit a simpler formulation in terms of gauge-theory building blocks. A systematic framework for finding this description is provided by the double-copy construction 
introduced by Bern, Carrasco, and one of the current authors \cite{BCJ,BCJ2}. The double copy applies to tree- and loop-level amplitudes~\cite{Bern:2012uf,Bern:2013uka,Johansson:2017bfl} as well as classical solutions~\cite{Classical,Goldberger:2016iau,Adamo:2017nia}, and extends earlier string-theory results by Kawai, Lewellen, and Tye \cite{KLT}.
Recent success in reformulating large families  of Maxwell-Einstein (ME) \cite{Carrasco:2012ca,Chiodaroli:2013upa,Johansson:2014zca,Chiodaroli:2015wal, Anastasiou:2016csv} and Yang-Mills-Einstein (YME) supergravities \cite{Chiodaroli:2014xia,Chiodaroli:2015rdg,Chiodaroli:2017ngp} in the double-copy language has prompted the proposal that all theories of gravity could be regarded as double copies of some sort \cite{Chiodaroli:2015wal} (see also \cite{Anastasiou:2017nsz}). Generalizing these constructions to \textit{gauged} ME and YME supergravities constitutes a major step towards  establishing this proposal, and has the potential for incorporating a large body of supergravity literature into the rapidly-developing field of amplitude calculations.

In this letter, we propose a general strategy for expressing gauged supergravities as double copies. The main result is that amplitudes with the correct properties can be obtained from 
those of a theory with spontaneously-broken gauge symmetry 
and a gauge theory with broken supersymmetry.   
We present an explicit example in which we gauge a $U(1)_{\text{R}}$ subgroup of the $SU(2)_{\text{R}}$ R-symmetry in theories belonging to the so-called generic Jordan family of $\cN=2$ ME supergravities. 

Gauged supergravity always involves a minimal coupling between (some of the) gravitini and one or more vector fields. Consequently, for Minkowski vacua, there exist  non-vanishing gravitini--vector amplitudes 
\begin{equation}
{\cal M}_3 \big(1 \overline{\psi}_i , 2\psi_j, 3 A^a \big) = i g_{\text{R}} t^a_{ij} \bar v_1^\mu \cancel{\varepsilon}_3 v_{2\mu} + {\cal O}(g_{\rm R}^0)
 \label{gravitinoamp} \, ,
\end{equation}  
where $g_{\text{R}}$ is the gauge coupling constant, $v_{l\mu} (l=1,2)$ are the gravitini's polarizations, and $t^a_{ij}$ 
are the representation matrices of the gauged R-symmetry subgroup, acting on the two gravitini. We omitted terms involving field strengths which do not explicitly depend on $g_{\rm R}$; these are unrelated to the gauging.  
While seemingly innocuous, the amplitude $(\ref{gravitinoamp})$ is not invariant under a linearized supersymmetry transformation, 
$v_{l\mu} \rightarrow v_{l\mu} + k_{l\mu} \epsilon$ 
(the spinor $\epsilon$ obeys $\cancel{k}_l \epsilon=0$ to preserve the gamma-tracelessness of $v_{l\mu}$).
Hence, assuming that the gauging procedure preserves the supersymmetry of the Lagrangian, the amplitude above must 
belong to a theory with spontaneously-broken supersymmetry (possibly partially). 
Since local supersymmetry can no longer be used to reduce the gravitino's physical polarizations down to two, a gravitino now has four distinct polarization states corresponding to a massive spin-3/2 particle. 
Thus, we need to consider a double-copy construction valid for massive gravitini.

\smallskip

{\bf Gauged supergravities as double copies:}
The double-copy construction of \cite{BCJ2} starts from gauge-theory amplitudes organized in terms of cubic graphs whose edges are labeled by representations of the gauge group. The color factor $c_i$ of each graph is obtained by dressing each vertex with the corresponding group-invariant symbol; the kinematic numerator $n_i$ of each graph includes the dependence on external polarizations as well as loop and internal momenta.
If (a) two gauge theories have common mass spectra and conjugate gauge-group representations (so that gravity states can be associated to gauge-invariant bilinears of gauge-theory states) and (b) the kinematic numerators ($n_i$ and $\tilde{n}_i$) obey the same algebraic identities as the color factors (manifesting color/kinematics  (C/K) duality) then 
\eq
\mathcal{M}^{(L)}_n \!= \!i^{L+1}\Big(\frac{\kappa}{2}\Big)^{2L+n-2} \!\!\!\! \sum_{i\in {\rm cubic}} \! \int \!
\frac{d^{LD}\ell}{(2\pi)^{LD}}\frac{1}{S_i}\frac{n_i\tilde{n}_i}{\prod_{\alpha_i} \! d_{\alpha_i}}
\label{Sugra}
\eqe
gives the corresponding $n$-point $L$-loop supergravity amplitude. Here $\kappa$ is the gravitational coupling, $S_i$ are symmetry factors, $1/d_{\alpha_i}$ are (possibly massive) propagators, and $D$ is the spacetime dimension.
For gauge-theory amplitudes which lack manifest C/K duality, generalized double-copy constructions have been proposed~\cite{GenDC}.

The freedom of choosing the two gauge theories is critical for having a double-copy description for large families of (super)gravities. Given the  number of explicit constructions to date, it has been  suggested that all gravity theories may have this property \cite{Chiodaroli:2015wal}. For our purpose, we must choose two gauge theories whose spectra and interactions allow for a factorized form of the amplitude in eq.~(\ref{gravitinoamp}). The product between a trilinear 
vector interaction and a minimally-coupled fermion--scalar interaction leads to the expected gravity states and interactions; the absence of explicit momenta in eq.~(\ref{gravitinoamp})
implies that the trilinear vector interaction has no derivatives and thus belongs to a spontaneously-broken gauge theory. Together with the
presence of massive gravitini, this implies that the fermions of the other gauge theory must be massive. 
We therefore propose that gauged supergravities around Minkowski vacua can be presented as double copies of a spontaneously-broken gauge theory and a gauge theory whose supersymmetry is explicitly broken by fermion masses. Schematically, the double copy is
  \begin{eqnarray}
\Big(\text{gauged SUGRA} \Big) = \Big(  {\text{Higgsed YM}} \Big) \otimes \Big(\text{s}\cancel{\text{upe}\vphantom{YM}}\text{r}   \  \text{YM}  \Big).\,  
 \end{eqnarray}


{\bf A simple realization:} 
To illustrate the proposed construction, we take as the left gauge theory (GT$_{\rm L}$) an $SU(N+M)$ 
YM-scalar theory with 4D Lagrangian
\begin{equation}
{\cal L}_{0} = {1 \over g^2} \Tr\Big[-\frac{1}{4} F_{\mu \nu}F^{\mu \nu} -\frac{1}{2} D_\mu\phi^a D^\mu\phi^a  +\frac{1}{4} [\phi^a, \phi^b]^2 \Big] ,
\end{equation}
with $a,b=1,\ldots, n$. As discussed above, the gauge symmetry is spontaneously broken; we choose a scalar VEV $\phi^a \rightarrow \phi^a + \langle\phi^a\rangle$ of the form
\be
\langle \phi^a \rangle = V^a \times \text{diag}\Big(\frac{1}{N}\id_N, -\frac{1}{M}\id_M\Big) , 
\label{vev1}
\ee
where $V^a$ is constant.
The subgroup $G=SU(M)\times SU(N)\times U(1)$ remains unbroken and the spectrum is
\begin{equation} 
\text{GT}_{\rm L}: \ \ 
\big\{ A_\pm, \  \phi^a \big\}_\cG \  \oplus   \  \big\{ W_{\hat \mu}, \ \varphi^s \big\}_\cR \  \oplus \  \big\{ \overline{W}_{\hat \mu} , \ \overline{\varphi}^s   \big\}_{\overline \cR}  \, , 
\end{equation} 
where ${\cal G}$ denotes the adjoint representations of $G$ and ${\cal R}$ and  ${\overline {\cal R}}$ are the bifundamental $(N, {\overline M})$ and $({\overline N}, M)$ representations.
All fields transforming in the $\cR, \overline \cR$ representations have the same mass $m$.
The index $s=2,\ldots, n$ runs over the massive scalars, while $\hat  \mu$ runs over the three physical polarizations of the massive $W$s. It was shown in ref.~\cite{Chiodaroli:2015rdg} that this theory obeys C/K duality. 

The right gauge theory (GT$_{\rm R}$) has explicitly-broken supersymmetry and Lagrangian
 \begin{eqnarray}
 {\cal L}_{\cN \! \cancel{\, =\, }  2} \!\!\! &=& \!\!\! {1 \over g^2}\Tr\Big[-\frac{1}{4}F_{\mu \nu}F^{\mu \nu} -\frac{1}{2}D_\mu \varphi_\alpha D^\mu \varphi_\alpha +\frac{1}{4} [\varphi_\alpha, \varphi_\beta]^2   \no \\
 && \hskip0.7cm
 \null +\frac{i}{2}{\overline\chi} \Gamma^\mu D_\mu \chi +  \frac{1}{2} {\overline\chi} \Gamma^\alpha [\varphi_\alpha + \langle \varphi_\alpha \rangle , \chi] \Big] \, ,
 \label{LNeq2}              
 \end{eqnarray}
where $\chi$ is a six-dimensional Weyl fermion and $\alpha,\beta=5,6$ (this compact notation reflects the six-dimensional origin of the theory).  
This theory preserves C/K duality because it can be realized as the orbifold of a spontaneously-broken pure ${\cal N}=2$ SYM theory. Indeed, 
we begin with the $SU(N+M)$ ${\cal N}=2$ SYM theory and spontaneously break the gauge group to  $G=SU(N)\times SU(M) \times U(1)$ 
by introducing  a VEV
\begin{equation}
\langle \varphi_\alpha \rangle = \widetilde V_\alpha \times \text{diag}\Big(\frac{1}{N}\id_N, -\frac{1}{M}\id_M\Big)  , \label{vev2}
\end{equation}
which is chosen to have the same magnitude as the one in the left 
gauge theory, $(V^a)^2 = (\widetilde V_\alpha)^2$, so that the two theories have common mass spectra. Conjugation by the matrix $\gamma = \text{diag}(\id_N, -\id_M)$ is a symmetry of the Lagrangian and so is the sign flip of fermion fields. We may therefore orbifold by their composition:
\be
A_\mu \mapsto \gamma A_\mu \gamma^{-1} \ ,
\quad
\chi\mapsto -  \gamma \chi \gamma^{-1} \ ,
\quad
\varphi\mapsto \gamma \varphi \gamma^{-1}
\ .
\label{orbifold}
\ee
Since, as shown in \cite{Chiodaroli:2013upa,Chiodaroli:2015rdg}, each of these operations preserves C/K duality, so must the resulting theory.
Its Lagrangian is that of eq.~(\ref{LNeq2}) and its spectrum is
\begin{eqnarray} 
\text{GT}_{\rm R}: \ \ \ 
\big\{ A_\pm,  \ \  \varphi_\alpha  \big\}_\cG \ \oplus \ \big\{ \chi \big\}_{\overline \cR}  \ \oplus \ \big\{ \overline \chi \big\}_{ \cR}  \ .
\end{eqnarray}

{\bf Gauging} $U(1)_{\text{R}}$ {\bf in $\cN=2$ supergravities:}
General ME supergravity theories with $\cN=2$ supersymmetry in five dimensions were constructed by Sierra, Townsend, and one of the current authors~\cite{Gunaydin:1983bi,GSK}. Their gaugings were studied in refs. \cite{Gunaydin:1984ak,Gunaydin:1984pf}; gaugings that require dualization of some of the vector fields to tensor fields were constructed later~\cite{GunaydinZagermann,Gunaydin:2000xk}.  Four-dimensional ME supergravities and their gaugings were studied in refs.~\cite{deWit:1983rz,deWit:1984pk,deWit:1984px,Cremmer:1984hj,Gunaydin:2005bf} (see~\cite{Freedman:2012zz} for further references).
The fields of 5D ME supergravity  with $n$ vector  multiplets are
\begin{equation} \text{MESG}: \ \ \ 
\{ e_{\mu}^{m}, \Psi_{\mu}^{i}, A_{\mu}^{I} ,
\lambda^{i a}, \varphi^{x}\} \ ,
\end{equation}
where $I = 0,1, \ldots, n$;  $a,x= 1,\ldots, n$, and $i,j=1,2$ are R-symmetry indices~\cite{Gunaydin:1983bi}. 
ME theories are completely specified by 
the cubic polynomial $\mathcal{V}(\xi^I)\equiv (2/3)^{3/2}C_{IJK}\xi^{I} \xi^{J} \xi^{K}$, where $\xi^I$ are coordinates of a $(n+1)$-dimensional ambient space and $C_{IJK}$ is a constant symmetric tensor. The scalar fields parametrize  the $\mathcal{V}(\xi)=1$ hypersurface in this ambient space. The metric $\text{\it \aa}_{IJ} $ of the kinetic energy term of the vector fields  is given by the restriction of  the ambient-space metric  to this hypersurface; it is written in terms of the vielbeine $(h_I , h_I^a)$  as $\text{\it \aa}_{IJ} = h_Ih_J + h^a_I h^a_J$ (see ref.~\cite{Gunaydin:1983bi} for explicit expressions). 
Thus, as is relevant for the amplitude perspective, theories in the ME class are uniquely specified by their spectra and three-point interactions.  

In this letter we will focus on the ME supergravities belonging to the generic Jordan family with symmetric target spaces in five and four spacetime dimensions. They 
have $n>1$ vector multiplets and are defined by the cubic polynomial 
${\cal V}(\xi) = \sqrt{2}  \xi^0 [ (\xi^1)^2 -  (\vec{\xi}\cdot\vec{\xi})]$. Their double-copy construction  was given in ref.~\cite{Chiodaroli:2014xia}.

As shown in refs.~\cite{Gunaydin:1984pf,Gunaydin:2000xk}, it is possible to gauge a $U(1)_{\text{R}}$ subgroup of the R-symmetry group $SU(2)_{\text{R}}$
for all ME supergravity theories. The resulting actions admit Minkowski vacua with spontaneously-broken supersymmetry.  Thus we expect them to admit a double-copy construction as explained  above. The relevant Lagrangians are obtained by covariantizing derivatives  on the fermions with respect to the $U(1)_{\text{R}}$ gauge field $ V_I A^I_\mu$  defined by an $(n+1)$-dimensional constant vector $V_I$,
\begin{eqnarray}
\mathcal{D}_{\mu}\Psi_{\nu}^{i}
&\equiv& \nabla_{\mu}\Psi_{\nu}^{i}+\gR V_{I}A_{\mu}^{I}\delta^{ij}
\Psi_{\nu j}\nonumber \ , \\
\mathcal{D}_{\mu}
\lambda^{i a}
&\equiv&  \nabla_{\mu}\lambda^{i a}+\gR V_{I}A_{\mu}^{I}\delta^{ij}
\lambda_{j}^{ a}   \ ,
\end{eqnarray}  
and adding the following terms to the 5D Lagrangian:
\begin{eqnarray}
\delta {\cal L} &=&-\frac{i\sqrt{6}}{8} \gR {\overline{\Psi}}_{\mu}^{i}
 \Gamma^{\mu\nu}
 \Psi_{\nu}^{j}\delta_{ij} P_{0}-\frac{1}{\sqrt{2}}
 \gR{\overline{\lambda}}^{i a}
 \Gamma^{\mu}
 \Psi_{\mu}^{j}\delta_{ij} P_{a}\nonumber\\
 &&\null +\frac{i}{2\sqrt{6}}\gR{\overline{\lambda}}^{ia}
 \lambda^{jb}\delta_{ij} P_{a b}-\gR^{2}P^{(\text{R})} \ .
\end{eqnarray}
The coefficient functions $P_0,P^a,$ and $P^{ab}$ 
are given in terms of  $V^I$ as
\begin{eqnarray}\label{U1cons}
P_{a}(\varphi)&=&\sqrt{2}h^{I}_a V_{I}\label{U1cons1} \no \ , \quad
P_{0}(\varphi) = 2 h^{I}V_{I}\label{U1cons2} \no \ , \\
P_{ab}(\varphi)&=&\frac{1}{2}\delta_{ab}P_{0}+2\sqrt{2} 
T_{abc}P^{c}\label{U1cons3} \ ,
\end{eqnarray}
with $T_{abc} = C_{IJK} \, h^I_a h^J_b h^K_c$.
The scalar potential $P^{(\text{R})}(\varphi)$ is given by \cite{Gunaydin:1984pf,Gunaydin:2000xk} 
\begin{equation}
P^{(\text{R})}(\varphi)=-(P_{0})^{2}+P_{a}P^{a} {=}  -4C^{IJK}V_IV_Jh_K \ ,
\end{equation}
where the indices of the constant tensor $C_{IJK}$ are raised by the inverse metric $\text{\it \aa}^{IJ} $.  For the generic Jordan family 
$C^{IJK}=C_{IJK}$. 

The deformation breaks the R-symmetry down to a $U(1)_{\text{R}}$ subgroup.
Minkowski vacua  correspond to gauging with vanishing potentials, $P^{(\text{R})}(\varphi)=0$; they break supersymmetry spontaneously \cite{Gunaydin:1984pf,Gunaydin:2000xk}. Up to rotations and overall rescaling, the simplest choice of $V_I$ leading to theories with this property is
$V^{(\pm)}_I  =   \big(0 , 1, \pm 1 , 0, \ldots , 0 \big)$ \cite{note}. This choice breaks the global symmetry group down to the Euclidean group $E_{(n-2)}$ for $(n-2)$ internal dimensions.  
To study the spectrum of the theory, it is convenient to redefine the massive gravitini as 
\begin{equation}
\xi^i_\mu = \Psi_\mu^i - {i \over \sqrt{12}} \Gamma_\mu \lambda^{ia} \frac{P_a}{P_0} + {\sqrt{2} \over \gR P_0 } {\cal D}_\mu \Big({ P_a  \lambda_i^a \over P_0} \Big) \ .
\end{equation}
After this operation, the Goldstino  field $\eta^i = \lambda^{ia} P_a/P_0$  no longer appears in the Lagrangian having been eaten by the gravitino which becomes massive (this is analogous 
to the unitary gauge in spontaneously-broken YM theories). 
Mass matrices for gravitini and remaining spin-1/2 fields can be written as
\begin{eqnarray}
M^\xi_{ij} =   {\sqrt{6} \over 4} \gR P_0  \delta_{ij} 
 , \quad
M^\lambda_{abij} = { \gR \over \sqrt{6}} \Big( \! P_{ab} \! - \! {5 \over 2} {P_a P_b \over P_0} \! \Big) \delta_{ij} . \ \
\end{eqnarray}
Taking into account that the non-vanishing coefficient matrices at the scalar base point are $P_0|_{c_I}=2 \sqrt{2 / 3}$, $P_1|_{c_I}=- \sqrt{2 / 3}=P_{11}|_{c_I}$, $P_2|_{c_I}=\mp \sqrt{2}$, $P_{12}|_{c_I}= \pm2\sqrt{2}$, $P_{22}|_{c_I}=\sqrt{6}$, and $P_{ss}|_{c_I}=-1$ ($s=3,\ldots,n$),
it is immediate to verify that the masses of the two gravitini and  one pair of  spin-1/2 fermions are  $m=\gR$. The remaining non-zero fermion masses are equal to $- \gR$.

A direct comparison of double-copy amplitudes with supergravity calculations requires that 
we properly identify the mass dependence (i.e. the dependence on $g_R$ in supergravity).  
Apart from its explicit appearance in the Lagrangian, 
in both gauge theory and supergravity the mass is also hidden in the massive particle wave functions.
To expose it we, shall use spinor-helicity notation and reduce the supergravity Lagrangian to four dimensions. 
For the 5D spinors rewritten as Dirac spinors, the reduction is straightforward. The reduction of a massive gravitino yields 
the 4D gravitino $\xi_\mu$ and a further spin-1/2 field $\xi$. 
The precise decomposition of the 5D gravitino is chosen such that the 4D quadratic terms are canonically normalized.


To obtain diagonal kinetic terms for the bosons in the  4D Lagrangian, we dualize the vector $A^{-1}_\mu$ from dimensional reduction of the 
graviton, and redefine fields as
\begin{eqnarray}
\left(\begin{array}{c} A^{-1}_\mu \\ A^{0}_\mu\\A^{1}_\mu \end{array} \right)
 &\rightarrow& 
-{1 \over 4}\left(\begin{array}{ccc}
-{1 } &  {1 } &     \sqrt{2}   \\
{2} &  -{2}  & {2 \sqrt{2}}  \\
 {2 \sqrt{2}} & {2 \sqrt{2}} & 0  \end{array}\right) \!\!\!
\left(\begin{array}{c} A^{-1}_\mu \\ A^{0}_\mu\\A^{1}_\mu \end{array} \right)\! . \qquad \label{redef}
\end{eqnarray}
After this operation, $A^{-1}_\mu$ is the 4D graviphoton and the vector identifying the $U(1)_{\text{R}}$ gauge boson is expressed as
\begin{equation}
V^{(\pm)}_A \! = \!   \big( V_{-1}, V^{(\pm)}_I \big) \! = \! \Big(\textstyle-{1 \over \sqrt{2}} , \textstyle -{1 \over \sqrt{2}} , 0, \pm 1 , 0, \ldots , 0 \Big)  . \label{choiceafter}
\end{equation}

Supergravity amplitudes can now be straightforwardly computed from the Lagrangian and matched with the ones from the double-copy method. We focus in particular on the amplitudes involving the $U(1)_{\text{R}}$ gauge field and two gravitini, which have the form in eq. (\ref{gravitinoamp}), where $t^a_{ij}$ is replaced by the identity (note that the polarization vector-spinors $v_{l\mu}$ need to be transverse and gamma-traceless).  
Such amplitudes can be reproduced 
with the following double-copy field map for the fermions:
\begin{eqnarray}
&&  \xi_\mu = W_\mu  \otimes \chi  ~  - ~ 
W_\nu \otimes \Big({\gamma_\mu \over 3}  -  {ip_\mu\over 3 m} \Big)\gamma^\nu \chi \ , \no \\   
&& 
\xi =  W_\nu \otimes \gamma^\nu \chi
\ , \qquad \quad
(U \lambda)^s =  \varphi^s \otimes \chi   \ .  \quad
\end{eqnarray}
The combination on the first line is manifestly transverse and gamma-traceless. $U$ is a unitary matrix diagonalizing the spin-$1/2$ mass terms and the index $s=2,3,\ldots, n$ runs over all spin-1/2 fields except the Goldstino. Since the $U(1)_{\text{R}}$ gauging affects only the fermionic terms in the Lagrangian, the double-copy origin of the vector fields will be the same as for the ungauged construction~\cite{Chiodaroli:2014xia}:
\bea 
A^{-1}_+  &=&  A_+  \otimes z  \,, \qquad z= (\varphi^6 + i \varphi^5)/\sqrt{2}  \, ,   \no \\
A^0_+  &=&  A_+  \otimes \bar z 
\, , \qquad
\pm i A^1_\pm   =    \phi^1  \otimes A_\pm   \, , \no \\
A^s_\pm  &=&   \phi^s  \otimes A_\pm  \, .
\eea
The factor of $i$ arises because the double copy is most naturally formulated in a symplectic frame with 
$SO(n)$ compact isometry, which differs from the one singled out by dimensional reduction by the dualization of one vector field. The gauge boson defined by  (\ref{choiceafter}) has the following  simple double-copy realization:
\begin{eqnarray}
A^{V^{(\pm)}}_+ &=& - A_+ \otimes {\varphi^6 } \,\, \pm \,\,  \phi^2 \otimes A_+   \ .
\end{eqnarray}

In order to match the amplitudes from the double copy with the ones from the supergravity Lagrangian, we employ the massive spinor-helicity formalism, writing massive momenta as 
$p_i = p_i^\perp- {m^2 \over 2 p_i \cdot q} q $.  Here $q$ is a reference momentum and $p_i^\perp, q$ are both massless.
Polarizations for massive spinors are written as
$v^t_+ =\big(  |i^\perp] , \  m {|q \rangle  / \langle i^\perp q \rangle} \big)$ and $
v^t_- = \big( m {|q ]  / [ i^\perp q ] } , \ |i^\perp\rangle  \big)$.
Explicit expressions for the massive-vector polarizations can be found in \cite{Craig:2011ws} (see also \cite{Arkani-Hamed:2017jhn}).
We consider massive gravitini with $\pm$ polarizations and rewrite selected gravitini--vector amplitudes as ($I = 0, \ldots, n$)
\begin{eqnarray}
& \!\!\! & \!\!\! M_3 \big( 1 \bar \xi_+ , 2 \xi_-, 3 A^{-1}_+ \big) =  -\sqrt{2} i  m \Omega {V_{-1}}  {\langle   2^\perp q \rangle \over \langle  1^\perp q \rangle} \ ,   \no \\
& \!\!\! & \!\!\! M_3 \big( 1 \bar \xi_+ , 2 \xi_-, 3 A^{I}_+ \big) =   \sqrt{2} i  m \Omega { V_I}  {[  1^\perp q] \over [  2^\perp q]}  \ . \qquad  \label{ampAa}
\end{eqnarray}
We note that, aside from the gravitino minimal coupling, the first amplitude has a contribution coming from a 
cubic interaction of the form $2 i  F^{-1}_{\pm \mu \nu} \bar \xi^\mu {\cal P}_\mp \xi^\nu$, where ${\cal P}_\pm$ denotes the chirality projector and $A^{-1}_\mu$ is the graviphoton. 
The overall factor of $\Omega=[3^\perp 1^\perp]^3/ ([1^\perp 2^\perp] [2^\perp  3^\perp ])$ is equal to the gauge-theory amplitude between two massive and one massless vectors. The $(n+2)$-dimensional vector $V_A$ defines the choice of $U(1)_{\text{R}}$ gauge vector and is given in (\ref{choiceafter}). This result matches the one from the double copy provided that the gauge theory VEVs are  
\begin{equation}
\widetilde V_\alpha= \big( 0, m \big) \ , \qquad
V^{(\pm)a}= \big( 0, \pm m, 0, \ldots , 0 \big) \ . \label{vevs} 
\end{equation}
The magnitude of the VEVs in the two gauge theories determines the supergravity parameter $g_{\text R}$ or, alternatively, the masses of gravitini and spin-1/2 fields.
Similarly, the direction of the gauge-theory VEVs is identified with the supergravity vector $V_A$ which defines the $U(1)_{\text R}$ gauge field.
From the point of view of the underlying gauge theories, the vanishing of the first entry in each VEV arises because the scalar fields $\phi^1$ and $ \varphi_5$ descend from the 5D gluons which have no VEV (the remaining zeros can be understood from the $SO(n-1)$ symmetry of $\phi^s$).

\smallskip

{\bf Conclusions:}
We have presented an explicit realization of an infinite family of supergravity theories with gauged $U(1)_{\text{R}}$ in the double-copy framework, and verified the construction by analyzing how the spectrum and three-point amplitudes are deformed by the parameter $g_{\rm R}$. For this family, the double copy automatically gives that all fermionic fields have the same mass, in agreement with the supergravity Lagrangian.  On general grounds, the double copy is expected to be robust beyond our explicit checks since the two gauge theories that enter the double-copy construction of the generic Jordan family admit massive deformations which preserve C/K duality and break supersymmetry. Hence, the double copy applied to any multiplicity and loop order should give sensible (diffeomorphism-invariant \cite{Chiodaroli:2017ngp}) amplitudes in supergravities with spontaneously-broken supersymmetry.

The double-copy construction given in this letter extends straightforwardly
to gauged YME supergravites by gauging a compact subgroup of the little group of $V_I$, which determines the $U(1)_{\text{R}}$ gauge field.  Our results may be generalized in several other directions,
such as including hypermultiplets and partial supersymmetry breaking, as well as extensions 
to  $\mathcal{N} \geq 4$ gauged supergravities with partial or complete supersymmetry breaking.  
The basic feature of the construction is expected to remain unchanged: combining a spontaneously-broken gauge theory with a theory with broken supersymmetry 
yields the amplitudes of a supergravity with massive gravitini (and hence spontaneously-broken supersymmetry). 
Finally, the construction outlined here opens the door to a complete classification of gauged supergravities with Minkowski vacua and broken supersymmetry that admit double-copy realizations. Understanding these theories may provide clues for extending the double-copy structure to vacua with an Anti-de-Sitter spacetime.  
    
\bigskip

We thank Alexandros Anastasiou for useful discussions on related topics.  
We are grateful to the Kavli Institute for Theoretical Physics for hospitality during the program ``Scattering Amplitudes and Beyond" where part of this work was completed. The research of M.C. and H.J. is supported in part by the Knut and Alice Wallenberg Foundation under grant
KAW 2013.0235, the Ragnar S\"oderberg Foundation under grant S1/16, and the Swedish Research Council
under grant 621-2014-5722. The research of R.R. was also supported in part by the US Department of Energy under
grant DE-SC0013699.


\begin{thebibliography}{99}
\bibitem{note}
In addition to $V_I^{(\pm)}$, the choice $V_I \! = \! ( \! \sqrt{2} , 0, \ldots , 0 )$ leads to flat gauging 
and will be considered in forthcoming work.
\end{thebibliography}
\end{document}